# Boulders on asteroid Toutatis as observed by Chang'e-2

Yun Jiang[1], Jianghui Ji[1*], Jiangchuan Huang[2], Simone Marchi[3], Yuan Li[4] & Wing-Huen Ip[4,5]

[1]Key Laboratory of Planetary Sciences, Purple Mountain Observatory, Chinese Academy of Sciences, Nanjing 210008, China, [2]China Academy of Space Technology, Beijing 100094, China, [3]Southwest Research Institute, Boulder, Colorado 80302, USA, [4]Space Science Institute, Macau University of Science and Technology, Taipa, Macau, [5]Institute of Astronomy, National Central University, Taoyuan, Taiwan. *E-mail: jijh@pmo.ac.cn.

Boulders are ubiquitously found on the surfaces of small rocky bodies in the inner solar system and their spatial and size distributions give insight into the geological evolution and collisional history of the parent bodies. Using images acquired by the Chang'e-2 spacecraft, more than 200 boulders have been identified over the imaged area of the near-Earth asteroid Toutatis. The cumulative boulder size frequency distribution (SFD) shows a steep slope of -4.4 ± 0.1, which is indicative of a high degree of fragmentation. Similar to Itokawa, Toutatis probably has a rubble-pile structure, as most boulders on its surface cannot solely be explained by impact cratering. The significantly steeper slope for Toutatis' boulder SFD compared to Itokawa may imply a different preservation state or diverse formation scenarios. In addition, the cumulative crater SFD has been used to estimate a surface crater retention age of approximately 1.6 ± 0.3 Gyr.

On 13 December 2012 at 8:29:58.7 UTC, Chang'e-2 successfully flew by the near-Earth asteroid 4179 Toutatis after accomplishing a lunar exploration phase and space-environment exploration at the Sun-Earth Lagrangian point. Chang'e-2's flyby with a distance of 770 ± 120 m (3σ) from Toutatis' surface ranks top for all known asteroid flyby missions, achieving a resolution better than 3 m/pixel for the closest image obtained at 18.3 km from Toutatis' surface[1]. Moreover, Chang'e-2's flyby allows us for the first time to closely observe the detailed morphology of Toutatis, although only ~ 45% of its surface was imaged due to the spacecraft's relatively high velocity of 10.73 km/s and the slow spin period of ~ 5.4 days for the asteroid[2-4]. The optical images acquired have clearly shown interesting geological features such as craters, boulders, lineaments and regolith, and an 800-meter-wide depression at the big end and the sharply perpendicular profile in the neck region[1, 5, 6]. Thus, Chang'e-2's expedition enables us to have an in-depth understanding of morphology and formation for Toutatis, providing ground truth for ground-based photometric, spectroscopic, and radar observations[7-11].

Boulder distributions and surface cratering give insight into geological evolution and collisional history of an asteroid. Comparative analyses of boulder SFDs on different asteroids can reveal processes of boulder formation and modification[12, 13]. Previous studies have shown that boulders on the surface of asteroids can be formed by impact cratering, or catastrophic disruption of the asteroid's parent body, or both, depending on the asteroid's size[13, 14]. Boulders on relatively large bodies (with an approximate mean diameter of 10–100 km), such as 21 Lutetia, 243 Ida and 433 Eros, are believed to be mostly produced by impact cratering, whereas boulders on much smaller bodies like Itokawa (with a mean diameter ~ 0.3 km) are likely to originate from the disruption of their larger parent bodies[13-17]. Toutatis, with an intermediate size (~ 4.6 × 2.3 × 1.9 km, ref. 1), lies within the two end-member cases described above. Several geological features suggest that Toutatis is bilobate and most possibly consists of a conglomeration of shattered fragments[1-3]. In this work, we identify boulders on Toutatis' surface and investigate their spatial and size distributions in order to constrain the formation and evolution of Toutatis.

## Spatial and size distribution of boulders

Considering the appearance and structure of Toutatis, we herein refer to the "head", "body" and "neck" to indicate the small lobe, large lobe and conjunction between them, respectively, following the descriptions of ref. 1. The spatial distribution of boulders on the imaged surface of Toutatis is shown in Fig. 1. A total of 222 boulders were identified over the imaged area (~ 6.68 km$^2$). They have sizes ranging from 10 to 61 m, with an average size of 22 m. The two largest boulders (> 50 m) are located in the neck region (Fig. 1). Approximately 90% of boulders are no larger than 30 m in diameter.

The cumulative size-frequency distribution (SFD) of boulders per unit area is shown in a log-log plot in Fig. 2. We limit our analysis to boulders larger than 20 m in size, which exhibits a slope (power-index) of -4.4 ± 0.1. In addition, we divide all boulders into two populations for the purpose of assessing the boulder distributions on the head and the body for comparison. By using the sharply perpendicular silhouette about the neck region as a boundary, over 50 boulders are counted on the head, and 160 rocks are identified on the body. Their separate distributions of boulders are normalized by the imaged areas of the head and body respectively, which give similar distribution trends with previous statistics[6]. Fig. 3 shows that the slopes of size distribution of boulders above 20 m are -3.9 ±0.2 and -4.5 ±0.1, respectively.

The number density of boulders larger than 20 m over the Toutatis' surface is ~ 17/km$^2$, which is a bit smaller than that of Itokawa (~ 30/km$^2$), but much larger than that of Eros (~ 2/km$^2$)[13,18]. The stark difference between Eros and Toutatis may result from the diverse formation process of boulders as discussed below. The number density of boulders larger than 20 m on the head (~ 15/km$^2$) is roughly the same as that on the body (~ 18/km$^2$), which bears a resemblance to that on the entire imaged side of Toutatis.

Toutatis' cumulative boulder SFD shows a slope of -4.4 ±0.1 for boulder sizes between 20 m and 60 m, which is significantly steeper than the slopes of -3.2 for Eros boulders with sizes ranging from 15 to 80 m, and of -3.3 ±0.1 for Itokawa boulders with sizes ranging from 6 to 38 m [13,18]. Other small objects like the Martian moon Phobos also reports a similarly shallower slope[19]. The slope of cumulative boulder SFD steeper than -2 is suggestive of very fragmented material, and a steeper slope is further indicative of greater degree of fragmentation[13,20]. Boulders could be comminuted by later impacts, which will steepen their size distribution because the largest sizes may be destroyed but not replenished[15]. The steep slope obtained for Toutatis implies that the boulders may have experienced much processing in the geological evolution, including breaking, sorting, and transporting[13,17]. As most boulders above 20 m are probably surviving fragments from Toutatis' parent body as discussed below, relatively steeper slope may suggest that the body (-4.5 ±0.1) has suffered more energetic impacts than the head (-3.9 ±0.2) does. If this is the case, the larger fracturing of the body may be one of reasons leading to less dense internal structure than the head, consistent with differences in the observed ratios of moments of inertia[21].

**Source of boulders**

As discussed, boulders are usually produced on asteroids by impact cratering or/and catastrophic disruption of the parent body. In either case, only those boulders ejected at velocities smaller than the escape velocity of the final body can be re-accreted on the asteroid's surface. As for the impact cratering process, ejected fragments are somewhat associated with the source crater[13].

Previous investigations revealed the correlations between the largest boulder size (L) and the source crater diameter (D) for the Moon, the Martian moons Phobos and Deimos, as well as the asteroid Ida[16, 22]. One of the empirical relations is expressed as $L \sim 0.25\ D^{0.7}$, with L and D in meters, which was applied for Eros and Itokawa[13, 23]. For Toutatis, the largest boulder is ~ 61 m in geometric mean diameter. According to the above empirical relationship, the corresponding source crater is ~ 2570 m, which is roughly four times larger than the largest impact crater (~ 530 m) on the surface of Toutatis[1]. Even for the depression at the big end of Toutatis, its size is only ~ 800 m. Due to the limited dimension of Toutatis, the probability of the existence of km-scale impact craters on the side not imaged by Chang'e-2 is very small. Moreover, the largest depression (possibly an impact crater) identified based on global radar-derived model was 750 m across[10]. Therefore, we infer that the largest boulder is not produced by the impact cratering. As a matter of fact, the maximum diameter of boulders generated by the largest impact crater (~ 530 m) on the imaged side of Toutatis is ~ 20 m. Also, there exist about 30 boulders with sizes ranging from 30 to 61 m discerned on the surface of Toutatis. For other small bodies, such as Phobos and Ida, boulders of this size range correspond to craters with 1–10 km diameter[22], which are involved in the impact cratering. These observations provide strong evidence that most boulders (larger than 20 m) are probably surviving fragments from the parent body of Toutatis, re-accreted after the initial disruption.

In addition to the analysis above, we quantitatively examine the origin of boulders on the surface of Toutatis by estimating a total volume of ejecta represented by the boulders and a total volume of ejecta represented by the craters on Toutatis. In this work, 222 boulders with sizes from 10 to 61 m give a total volume $\sim 1.8 \times 10^6\ m^3$, on the basis of the assumption[23] that the height of a boulder is half of its geometric mean dimension. There are fifty craters with diameters from 40 to 530 m identified from the flyby images of Toutatis, as mentioned by ref. 1. According to similar estimation method for asteroids Eros and Itokawa[13, 23], the volume of materials produced in the formation of these craters can be estimated to be $\sim 1.9 \times 10^7\ m^3$, if the excavated volume is half that of the crater volume which is expressed as $V \sim 0.07 D^3$, where D is crater diameter in meters. Thus, the boulders are about 10% the excavated volume of these craters, which would become higher if we take all boulders < 20 m into account which are counted incompletely in this work. The percentage on Toutatis (10%) is much larger than that on Eros (0.4%) where most boulders are believed to originate from the Shoemaker crater, but smaller than that on Itokawa (25%) where most boulders are of endogenic origin, i.e., fragments from the parent body[13, 14]. The high ratio of the total volume of the boulders to the total excavated volume of the craters on Toutatis may also imply that most boulders cannot solely be created as products of cratering on the surface of Toutatis.

**Surface age**

Although optical images of Toutatis obtained by the Chang'e-2 spacecraft have a significantly better resolution than that of the previously published radar-derived model[10, 21, 24], the overall quality is relatively poor because they were acquired by the engineering camera, making the detection of degraded craters a herculean task. As a result, fifty craters with diameters from 40 to 530 m have been discerned and catalogued as impact craters with various degree of confidence[1]. Toutatis has an overall shallow slope of the cumulative crater SFD (Fig. 4), with a distinct shallower slope for diameter smaller than ~ 0.15 km. Note that this size is well above the spatial resolution, therefore the slope turnover is unlikely due to limited resolution. The observed roll over of the crater SFD at smaller sizes shown by Toutatis qualitatively resembles that of Stein crater SFD, also shown in Figure 4 for a comparison. As suggested for Steins[25], the rollover of the crater SFD at small diameter may indicate the presence of some crater obliteration mechanisms. Small craters can be erased by regolith displacement due to the cumulative effects of seismic shaking of small impacts or a large impulse of erasure triggered by the formation of large craters. Similar processes have been demonstrated to be common on small bodies like Steins, Ida, Eros, and Itokawa[25-28].

In order to evaluate the crater retention age of Toutatis we estimate the impactor flux and convert impactor size to crater size by adopting a crater scaling law[25, 29]. Given the likely highly fractured nature of Toutatis, we adopt a rubbly material crater scaling law, with the impact flux following the model for the dynamical evolution of asteroid belt[30]. As for the impact rate, we choose average impact condition for main belt asteroids. This is mainly because the time spent by Toutatis in the main belt is typically 100 times larger than the time spent on the near-Earth orbit, hence most of the observed craters on Toutatis are formed via collisions within the main belt. In order to estimate the crater retention age we fit the crater SFD at sizes > 0.15 km (Fig 4), the resulting surface age of Toutatis is estimated to be ~ 1.6 Gyr, for rubbly material and a strength of $2 \times 10^6$ Pa, supporting the validity of our assumption[31]. The nominal age is relatively close to an age estimate for Steins (using a similar crater scaling law). We warn, however, that low number statistics (particularly at large crater sizes) preclude a robust age assessment for both asteroids. Moreover, the cratering age for a small asteroid is a function of its material strength, and weaker materials would result in younger ages. The formal error estimate for our age estimate is ±0.3 Gyr, but in reality it is larger due to the uncertain factors described above.

## Discussion

Boulders are ubiquitous on the surfaces of asteroids. For a relatively large asteroid like Lutetia, gravity is strong enough that boulders ejected from impacts are concentrated around the source crater. For small asteroids like Itokawa, most of ejected boulders must have been lost due to the low escape velocity, and those re-accreted are likely to be scattered globally on the surface. As far as the present investigation is concerned, most of boulders on the surface of asteroids appear to be connected to source craters, like the Shoemaker crater on Eros, the Stickney crater on Phobos and the central crater on Lutetia, whereas the distribution of observed boulders on Itokawa likely results from processes involved in Itokawa's re-accretion following the disruption of the parent body[13, 15, 23, 32].

Toutatis can be regarded as an intermediate body between Eros and Itokawa in many respects, for instance, body size, porosity, as well as the number density of boulders > 20 m and the ratio of the total volume of the boulders to the total excavated volume of the craters as mentioned above. As small-sized objects, Toutatis and Itokawa show a relatively uniform distribution of boulders, without local concentration around craters. The recent research shows that the size distribution and segregation of boulders may be correlated with the gravity field[33].

Boulders could also have been disintegrated by subsequent micrometeoroid impacts and thermal fatigue processes[12, 34]. Thermal fragmentation induced by the diurnal temperature variations is considered to be a dominant mechanism of rock weathering and fragmentation on kilometer-sized asteroids[12]. Toutatis' boulders exhibit a clearly steeper slope than Itokawa, probably implying a different preservation state of boulders. If so, the preservation state of boulders may have something to do with the past orbital evolution. As discussed by ref. 35, near-Earth objects have a rather complex orbital evolution, with significant oscillations of the perihelion distance. The latter is important because the closer to the Sun the object is, the higher the thermal effects. Interestingly, numerical simulations[35] show that Toutatis minimum perihelion could have been as low as 0.02 AU with respect to 0.06 AU of Itokawa, suggesting a higher relevance of thermal effects for Toutatis.

## Methods

We define boulders as rocks and features with (1) a clearly identifiable brightness variation and (2) a bright positive relief with shadow next to it, following the criteria in the identification of boulders on the Lutetia surface[15]. In order to investigate the size-frequency statistics of boulders on Toutatis, measurements were made mainly based on seven optical images with resolutions of ~ 3.6–8.3 m/pixel and imaging distances of 29.0–67.7 km[1]. When calculating the size of boulders, different approaches have been used. Ref. 17 took the maximum size as the geometrical diameter of a boulder. Afterwards, ref. 23 took the geometric mean of the long and short axes of a boulder. Recently, ref. 18 used a boulder diameter more representative of its mass or volume. The difference among these methods is owing to the deviations in definition of boulder size.

We cannot simply apply any of above methods to our measurements, as we have not connected the shape model of Toutatis with the images acquired by Chang'e-2. Furthermore, some facets of Toutatis imaged by Chang'e-2 are not perpendicular to camera shooting direction. In our work, we choose to use the maximum length of the pixel line as the size for boulders resembling spherical shapes. For irregularly shaped boulders, we determine their size as geometric mean of maximum size and minimum size. The real size of boulders is estimated by evaluating width of pixels defining each boulder. It should be noted that the largest boulder has a smaller size than that previously measured[1, 6]. This is due to the difference in the measurements of boulder size. For the same boulder, the geometric mean in this work is smaller than the maximum in ref. 1.

1. Huang, J. C. *et al*. The Ginger-shaped Asteroid 4179 Toutatis: New observations from a successful flyby of Chang'e-2. *Scientific Reports* **3,** 3411 (2013).

2. Hudson, R. S. & Ostro, S. J. Shape and non-principal axis spin state of asteroid 4179 Toutatis. *Science* **270,** 84–86 (1995).

3. Ostro, S. J. *et al*. Asteroid 4179 Toutatis: 1996 radar observations. *Icarus* **137,** 122–139 (1999).

4. Zhao, Y. H. *et al*. Orientation and rotational parameters of asteroid 4179 Toutatis: New insights from Chang'e-2's close flyby. *MNRAS* **450,** 3620–3632 (2015).

5. Zhu, M. H. *et al*. Morphology of asteroid (4179) Toutatis as imaged by Chang'E-2 spacecraft. *Geophys. Res. Lett.* **41,** 328–333 (2014).

6. Zou, X. *et al*. The preliminary analysis of the 4179 Toutatis snapshots of the Chang'E-2 flyby. *Icarus* **229,** 348–354 (2014).

7. Hudson, R. S. & Ostro, S. J. Photometric properties of Asteroid 4179 Toutatis from light curves and a radar-derived physical model. *Icarus* **135,** 451–457 (1998).

8. Scheeres, D. J., Ostro, S. J., Hudson, R. S., DeJong, E. M. & Suzuki, S. Dynamics of orbits close to Asteroid 4179 Toutatis. *Icarus* **132,** 53–79 (1998).

9. Reddy, V. *et al*. Composition of near-earth asteroid (4179) Toutatis. *Icarus* **221,** 1177–1179 (2012).

10. Hudson, R. S., Ostro, S. J. & Scheeres, D. J. High-resolution model of asteroid 4179 Toutatis, *Icarus* **161,** 346–355 (2003).

11. Ostro, S. J. *et al*. Radar images of asteroid 4179 Toutatis. *Science* **270,** 80–83 (1995).

12. Delbo, M. *et al.* Thermal fatigue as the origin of regolith on small asteroids. *Nature* **508**, 233–236 (2014).

13. Thomas, P. C., Veverka, J., Robinson, M. S. & Murchie, S. Shoemaker crater as the source of most ejecta blocks on the Asteroid 433 Eros. *Nature* **413,** 394–396 (2001).

14. Fujiwara, A. *et al*. The rubble-pile asteroid Itokawa as observed by Hayabusa. *Science* **312,** 1330–1334 (2006).

15. Küppers, M. *et al*. Boulders on Lutetia. *Planet. Space. Sci.* **66,** 71–78 (2012).

16. Lee, P. *et al*. Ejecta blocks on 243 Ida and on other asteroids. *Icarus* **120,** 87–105 (1996).

17. Saito, J. *et al*. Detailed images of Asteroid 25143 Itokawa from Hayabusa. *Science* **312,** 1341–1344 (2006).

18. Mazrouei, S., Daly, M.G., Barnouin, O. S., Ernst, C. M. & DeSouza, I. Block distributions on Itokawa. *Icarus* **229,**181–189 (2014).

19. Thomas, P. C. *et al*. Phobos: regolith and ejecta blocks investigated with Mars Orbiter Camera images. *J. Geophys. Res*. **105,** 15091–15106 (2000).

20. Hartmann, W. K. Terrestrial, lunar, and interplanetary rock fragmentation. *Icarus* **10,** 201–213(1969).

21. Busch, M. W. *et al.* Internal structure of 4179 Toutatis. 2012 AGU Fall Meeting, P31A–1873 (2012).

22. Lee, S. W., Thomas, P. & Veverka, J. Phobos, Deimos, and the Moon: Size and distribution of crater ejecta blocks. *Icarus* **68,** 77–86 (1986).

23. Michikami, T. *et al*. Size-frequency statistics of boulders on global surface of Asteroid 25143 Itokawa. *Earth Planets Space* **60,** 13–20 (2008).

24. Takahashi, Y., Busch, M. W. & Scheeres, D. J. Spin state and moment of inertia characterization of 4179 Toutatis. *Astron. J.* **146,** 95 (2013).

25. Marchi, S. *et al*. The cratering history of asteroid (2867) Steins. *Planet. Space. Sci.* **58,** 1116–1123 (2010).

26. Asphaug, E. *et al*. Mechanical and geological effects of impact cratering on Ida. *Icarus* **120,** 158–184 (1996).

27. Richardson, J. E., Melosh, H. J. & Greenberg, R. Impact-induced seismic activity on asteroid 433 Eros: A surface modification process. *Science* **306,** 1526–1529 (2004).

28. Michel, P., O'Brien, D. P., Abe, S. & Hirata, N. Itokawa's cratering record as observed by Hayabusa: Implications for its age and collisional history. *Icarus* **200,** 503–513 (2009).

29. Barucci, M. A., Fulchignoni, M., Ji, J. H., Marchi, S. & Thomas, N. The fly-bys of asteroids 2867 Steins, 21 Lutetia, and 4179 Toutatis, Asteroid IV proceeding, in press (2015).

30. Marchi, S., Chapman, C. R., Barnouin, O. S., Richardson, J. E. & Vincent, J.-B. Cratering on Asteroids. Asteroid IV proceeding, in press (2015).

31. Marchi, S., Paolicchi, P., Lazzarin, M. & Magrin, S. A general spectral slope-exposure relation for S-type main belt and near-earth asteroids. *Astron. J.* **131,** 1138–1141 (2006).

32. Thomas, P. C. Surface features of Phobos and Deimos. *Icarus* **4,** 223–243 (1979).

33. Tancredi, G., Roland, S. & Bruzzone, S. Distribution of boulders and the gravity potential on asteroid Itokawa. *Icarus* **247,** 279–290 (2015).

34. Dombard, A. J., Barnouin, O. S., Prockter, L. M. & Thomas, P. C. Boulders and ponds on the Asteroid 433 Eros. *Icarus* **210,** 713–721 (2010).

35. Marchi, S., Delbo, M., Morbidelli, A., Paolicchi, P., Lazzarin, M. Heating of near-Earth objects and meteoroids due to close approaches to the Sun. *MNRAS* **400,** 147–153 (2009).

36. Holsapple, K. A., Housen, K. R. A crater and its ejecta: An interpretation of deep impact. *Icarus* **187,** 345–356 (2007).

37. Bottke, W. F. *et al*. The fossilized size distribution of the main asteroid belt. *Icarus* **175,** 111–140 (2005).

## Acknowledgments

This work is financially supported by National Natural Science Foundation of China (Grants No. 41403056, 11273068, 11473073), the Strategic Priority Research Program-The Emergence of Cosmological Structures of the Chinese Academy of Sciences (Grant No. XDB09000000), the innovative and interdisciplinary program by CAS (Grant No. KJZD-EW-Z001), the Natural Science Foundation of Jiangsu Province (Grant No. BK20131040, BK20141509), and the Foundation of Minor Planets of Purple Mountain Observatory.

## Author contributions

Y. J., J. H. J. and S. M. wrote the manuscript. J. H. J., J. C. H. and W. H. I. designed the research. S. M., W. H. I., and J. H. J. contributed ideas, discussions and comments on the manuscript. Y. J. and Y. L. analyzed the data. All authors contributed to the manuscript.

## Additional information

**Competing financial interests:** The authors declare no competing financial interests.

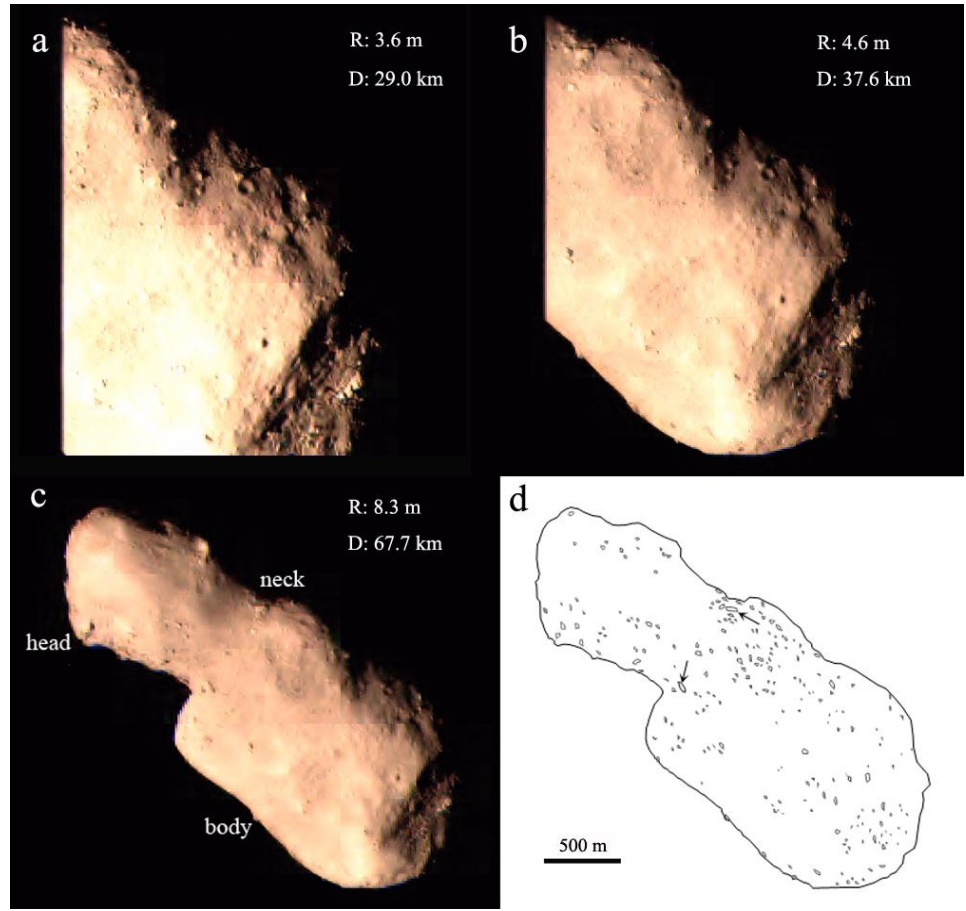

**Figure 1 | Outbound images display boulders on the surface of Toutatis (from a to c).** The left side of Toutatis is sheltered by the solar panel (a and b), whereas c is the first panoramic image acquired by Chang'e-2. The resolution (R) and imaging distance (D) are shown. The Sun-Toutatis-Chang'e-2 phase angle for outbound imaging is ~ 36.4°. (d) The sketch map shows the outline of Toutatis, and the approximate positions of 222 boulders identified in this work. The two largest boulders with geometric mean sizes larger than 50 m are marked by black arrows around the neck region.

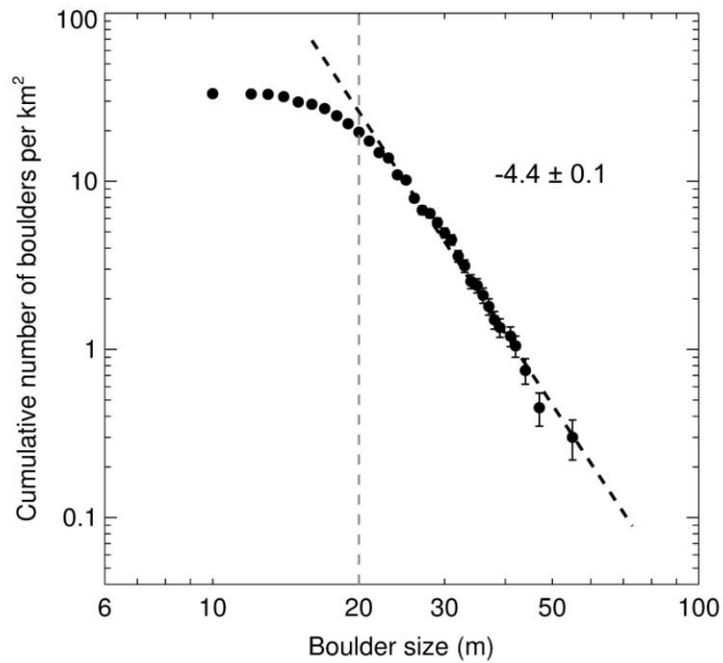

**Figure 2 | The cumulative size-frequency distribution of boulders per unit area on the entire imaged surface of Toutatis.** The horizontal axis is the boulder size defined as the geometric mean of maximum and minimum size of the boulder. The vertical axis is the number of boulders larger than a given size per unit surface. Vertical error bar indicates the square root of the cumulative number of counting boulders (Poisson statistics error) divided by the area. The count may be incomplete for boulders < 20 m, causing the distribution rolls over at smaller sizes. The vertical dashed line shows the location of 20 m boulders above which we limit our power law fit. The resulting power-law index of the best fit is -4.4 ± 0.1 for boulders > 20 m.

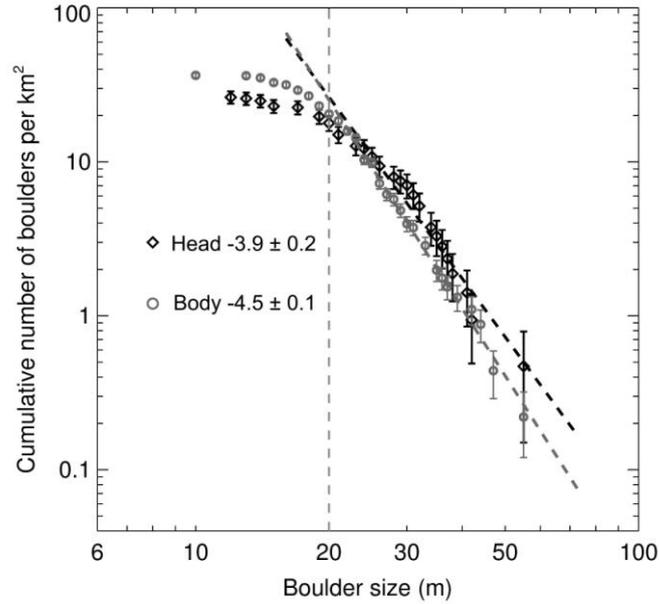

**Figure 3 | The cumulative SFD of boulders per unit area on the head and body of Toutatis.** Over 50 boulders are measured on the head, whereas 160 boulders are made out on the body, and their separate distributions of boulders are normalized by the areas of the head and body respectively. Their slopes for boulders > 20 m are -3.9 ± 0.2 and -4.5 ± 0.1, respectively.

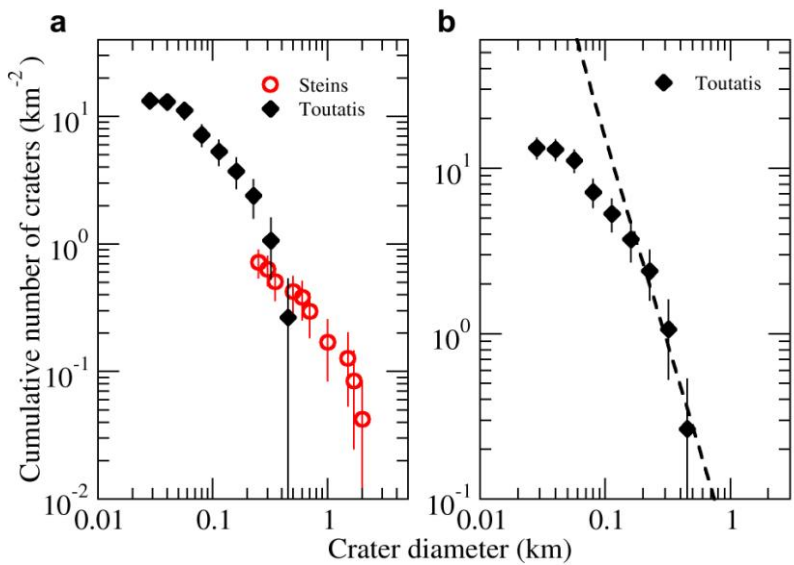

**Figure 4 | The size frequency distributions of craters on Toutatis.** (a) All identified craters on the imaged surface. For a comparison, the Steins cumulative SFD is also shown[25]. Error bars are estimated on the basis of poisson statistics of counts. (b) The best fit of cumulative crater SFD infers a cratering age of ~ 1.6 ± 0.3 Gyr, for rubbly material and a target strength of $2 \times 10^6$ Pa. Other parameters adopted for the age estimates are: target density (2.1 g/cm$^3$), impactor density (2.6 g/cm$^3$), and intrinsic collisional probability $2.86 \times 10^{-18}$ km$^{-2}$ yr$^{-1}$ corresponding to the average main belt value. The rubbly material crater scaling law is derived from ref. 36. The main belt impactor SFD is from ref. 37.